\pdfoutput=1
\documentclass[11pt]{article}
\usepackage[final]{acl}
\usepackage{times}
\usepackage{latexsym}
\usepackage[T1]{fontenc}
\usepackage[utf8]{inputenc}
\usepackage{microtype}
\usepackage{graphicx}
\usepackage{amsmath,amssymb}
\usepackage{booktabs}
\usepackage{multirow}
\usepackage{xcolor}

\title{LLM Latent Edge Measurement: Point-in-Time Economic Graphs\\
for Quantitative Investing from Corporate Disclosures}

\author{Fan Yang \\
  VinePeak Research \\
  \texttt{fyang@vinepeak.com}
  \And
  Lin Zhang \\
  VinePeak Research \\
  \texttt{linzhang@vinepeak.com}}

\begin{document}
\maketitle

\begin{abstract}
Standard industry classifications such as GICS assign each firm to exactly one
sector, yet the economic ties that transmit shocks between firms---supply
agreements, customer concentration, IP licensing, cloud and power
contracts---routinely cross sector boundaries and are disclosed only in
unstructured text. We treat the construction of a firm-level adjacency matrix
as a \emph{measurement} problem and propose an LLM-based pipeline that reads
corporate disclosures and outputs a weighted, directed, point-in-time network.
The pipeline combines (i) multi-agent evidence collection, in which
independent filing-reading and background-knowledge agents emit calibrated
relationship records; (ii) entity resolution of both named and \emph{latent}
(unnamed but inferable) counterparties; (iii) logit-pooled fusion of agent
votes; (iv) an inverse-document-frequency weighting that down-weights
ubiquitous counterparties (``economic stop-words'' such as TSMC or Samsung);
and (v) a double-sided relative threshold, borrowed from international-trade
network construction, that forms an edge whenever either firm is an important
partner of the other. Applied to 42 Nasdaq-100 constituents using their most
recent 10-K/20-F filings, the pipeline produces a 149-edge network in which an
adversarial audit confirms 88\% of sampled edges with weight $\ge 0.1$
(100\% including plausible-but-weakly-documented ties), and all refuted edges
concentrate in the lowest-weight tail. The measured network agrees with GICS
where it should (1.9$\times$ within-sector lift) while recovering
economically real cross-sector edges that GICS cannot represent, such as
nuclear power purchase agreements linking a utility to hyperscalers and the
GPU-cloud dependencies of an AI-infrastructure ecosystem. Ablations show that
each component---multi-agent fusion, IDF filtering, and relative
thresholding---contributes measurably to edge quality.\footnote{Code, extraction
records, and the measured matrices are released with the paper.}
\end{abstract}

\section{Introduction}

A large fraction of quantitative finance treats the cross-section of firms as
a set of independent observations grouped by an industry label. Yet firms are
nodes in a web of contracts: NVIDIA's data-center revenue is a function of
Microsoft's capital expenditure; Micron's equipment spending flows to ASML and
Lam Research; Constellation Energy's earnings now depend on twenty-year power
purchase agreements with hyperscalers. When information arrives at one node it
propagates along these edges with measurable delay
\citep{cohen2008economic,menzly2010market}, and portfolios sorted on linked-firm
returns earn predictable spreads precisely because the links are hard to
observe. The binding constraint is not the econometrics of spillover but the
\emph{measurement of the graph itself}.

Existing sources of firm networks each have a blind spot. Commercial
supply-chain databases (FactSet Revere, Bloomberg SPLC) rely on analyst
curation of the same disclosures we study, with documented coverage and
disclosure biases \citep{culot2023using}. Text-similarity networks such as
TNIC \citep{hoberg2016text} capture product-market relatedness but not the
direction or type of an economic dependency: two firms can sell similar
products while one is also the other's largest customer. Regulation S-K
customer-concentration disclosures \citep{cohen2008economic} are precise but
sparse---firms must disclose customers above 10\% of revenue, and may do so
without naming them (``Customer~A represented 17\% of revenue''). The
counterparty is then \emph{latent}: absent from any structured field, yet
frequently inferable from context by a reader with market knowledge.

This paper argues that large language models are well suited to exactly this
measurement task, and makes the task---not community detection, not return
prediction---the object of study. We ask: \emph{can an LLM-based pipeline
convert unstructured, partially anonymized corporate disclosure into a
well-calibrated adjacency matrix?} Three design ideas, adapted from network
construction in other domains, do most of the work.

\paragraph{Relative, double-sided importance.} International-trade networks
face extreme size heterogeneity: an absolute trade-volume threshold connects
every country to the United States and nothing to small open economies. The
standard remedy is a relative threshold---connect $i$ and $j$ if $j$ is a top
trading partner \emph{of $i$}, or vice versa. Firm networks have the same
pathology (Apple's smallest disclosed supplier relationship can exceed the
entire revenue of a mid-cap firm), and we import the same fix
(\S\ref{sec:threshold}).

\paragraph{Ubiquity filtering.} In legislative co-sponsorship networks, bills
with very many co-sponsors carry little information about latent political
alignment and are excluded before network construction. The corporate
analogue: a tie to TSMC, Samsung, or a hyperscaler cloud is so widespread that
sharing it says little about whether two firms belong to the same economic
community. We formalize this as an inverse-document-frequency weight on shared
counterparties (\S\ref{sec:idf}).

\paragraph{Measurement with uncertainty.} A single LLM asked ``are firms $i$
and $j$ related?'' produces a point answer with unknown reliability. We
instead run independent evidence agents with different information
sets---filing readers constrained to quote verbatim evidence, and background-knowledge
agents prompted for calibrated priors---and fuse their votes by weighted logit
pooling around a sparse prior (\S\ref{sec:fusion}), preserving edge-level
uncertainty in the final matrix.

We instantiate the pipeline on 42 constituents of the Nasdaq-100 (the universe
of the Invesco QQQ trust), chosen to span an AI-infrastructure ecosystem
(semiconductors, semi equipment, EDA, hyperscalers, AI cloud, power) plus
deliberately distant control groups (consumer staples, biotech). All filing
evidence is drawn from each firm's most recent annual report on SEC EDGAR,
giving a point-in-time snapshot as of each filing date.

Our contributions: (1) a formalization of LLM latent edge measurement,
including explicit handling of unnamed counterparties and evidence
provenance; (2) a multi-agent fusion scheme with entity-IDF filtering and
double-sided relative thresholds; (3) an empirical demonstration on real
EDGAR filings with an adversarial precision audit, GICS comparison, and
component ablations; and (4) a released dataset of $\sim$950 relationship
records with verbatim evidence quotes and the resulting matrices.

\section{Related Work}

\paragraph{Economic links and spillover.} \citet{cohen2008economic} use
mandated customer disclosures to show customer--supplier return
predictability; \citet{menzly2010market} find the same along input--output
links, and \citet{ali2020shared} show analyst co-coverage subsumes many
link-momentum effects. All take the graph as given; our contribution is
upstream of theirs.

\paragraph{Firm networks from text.} \citet{hoberg2010product,hoberg2016text}
build product-similarity networks from 10-K text; \citet{lee2015search} use
EDGAR co-search; \citet{kaustia2021common} use shared analysts;
\citet{schwenkler2019network} extract links from news co-mentions.
\citet{vamvourellis2023company} show LLM embeddings recover industry structure
from business descriptions. These capture relatedness; none measures typed,
directed dependencies with evidence.

\paragraph{Relation extraction in finance.} Supervised financial RE datasets
include FinRED \citep{sharma2022finred} and REFinD, built specifically over
SEC filings \citep{kaur2023refind}; \citet{rajpoot2023gptfinre} show in-context
LLM extraction is competitive. Supply-chain extraction from news and public
sources has moved from feature-based deep models \citep{wichmann2020extracting}
to LLM and RAG pipelines
\citep{liu2024supply,almahri2024enhancing,jackson2025supply}. Closest to us,
FinDKG \citep{li2024findkg} and FinReflectKG \citep{arun2025finreflectkg} build
financial knowledge graphs from news and S\&P-100 filings with agentic LLM
pipelines, and \citet{chen2023chatgpt} let ChatGPT emit a firm graph for GNN
stock prediction, while \citet{chung2023modeling} discover links from documents
for momentum spillover. Relative to this line we add: latent (unnamed)
counterparty resolution as a first-class problem, multi-agent probabilistic
fusion rather than single-pass extraction, ubiquity filtering, and
size-adjusted thresholding---together with an audit protocol that measures
whether the resulting \emph{matrix} (not just individual triples) is good.

\paragraph{Graphs for stock prediction and network model selection.} GNN stock
models consume firm graphs whose provenance strongly affects accuracy
\citep{chen2018incorporating,feng2019temporal,kim2019hats}; a measured,
documented adjacency matrix is a better input than a scraped one. Downstream
of measurement, block models \citep{holland1983stochastic} and network
cross-validation \citep{chen2018network,li2020network,wang2017likelihood}
provide the model-selection machinery for the community-detection questions we
defer to companion work.

\section{Problem Formulation}
\label{sec:problem}

Let $V$ be a universe of $N$ firms. At time $t$ we seek a weighted graph
$G_t=(V,E_t,W_t)$ where $W_{ij,t}\ge 0$ measures the strength of the economic
relationship between $i$ and $j$ using only information available by $t$. The
primitive object is not the edge but the \emph{directed dependency}
\begin{equation}
d_{i\to j,t}\;\in\;[0,1],
\end{equation}
the economic importance of counterparty $j$ to firm $i$ (share of revenue,
criticality of supply, licensing dependence). Dependencies are asymmetric by
construction: a supplier earning 40\% of its revenue from Apple has
$d_{\text{supplier}\to\text{AAPL}}$ high while Apple's reverse exposure may be
negligible.

The measurement input is a corpus $\mathcal{D}_t=\{D_{1},\dots,D_{N}\}$ of
firm disclosures (we use annual reports; the framework extends to transcripts
and news). Disclosure text names some counterparties explicitly
(``\emph{Three Mile Island Unit~1 \ldots supported by a 20-year power purchase
agreement with Microsoft}''), leaves others latent
(``\emph{one customer accounted for approximately 17\% of revenue}''), and
describes many exposures only at the category level (``\emph{we depend on a
limited number of foundries}''). An edge-measurement system must (i) resolve
entities across these levels, flagging which resolutions are inferred rather
than stated; (ii) attach type, direction, strength, and confidence to each
relationship; and (iii) preserve provenance so that any edge can be audited
back to text.

\section{Method}
\label{sec:method}

\begin{figure}[t]
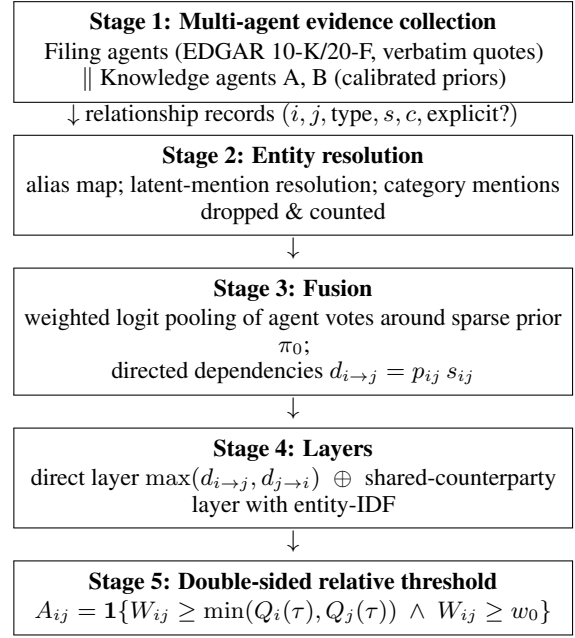

\centering
\footnotesize
\setlength{\fboxsep}{4pt}
\begin{tabular}{@{}c@{}}
\fbox{\parbox{0.92\columnwidth}{\centering
\textbf{Stage 1: Multi-agent evidence collection}\\[1pt]
Filing agents (EDGAR 10-K/20-F, verbatim quotes)\\
$\parallel$ Knowledge agents A, B (calibrated priors)}}\\[2pt]
$\downarrow$ relationship records $(i,j,\text{type},s,c,\text{explicit?})$\\[2pt]
\fbox{\parbox{0.92\columnwidth}{\centering
\textbf{Stage 2: Entity resolution}\\[1pt]
alias map; latent-mention resolution; category mentions dropped \& counted}}\\[2pt]
$\downarrow$\\[2pt]
\fbox{\parbox{0.92\columnwidth}{\centering
\textbf{Stage 3: Fusion} \\[1pt]
weighted logit pooling of agent votes around sparse prior $\pi_0$;\\
directed dependencies $d_{i\to j}=p_{ij}\,s_{ij}$}}\\[2pt]
$\downarrow$\\[2pt]
\fbox{\parbox{0.92\columnwidth}{\centering
\textbf{Stage 4: Layers}\\[1pt]
direct layer $\max(d_{i\to j},d_{j\to i})$\; $\oplus$\;
shared-counterparty layer with entity-IDF}}\\[2pt]
$\downarrow$\\[2pt]
\fbox{\parbox{0.92\columnwidth}{\centering
\textbf{Stage 5: Double-sided relative threshold}\\[1pt]
$A_{ij}=\mathbf{1}\{W_{ij}\ge \min(Q_i(\tau),Q_j(\tau))\ \wedge\ W_{ij}\ge w_0\}$}}
\end{tabular}
\caption{The latent edge measurement pipeline.}
\label{fig:pipeline}
\end{figure}

Figure~\ref{fig:pipeline} summarizes the pipeline; we describe each stage.

\subsection{Multi-agent evidence collection}
\label{sec:agents}

We deploy two types of evidence agents, instantiated with the same underlying
LLM but different information sets and instructions.

\textbf{Filing agents} retrieve each firm's most recent annual report from
EDGAR and emit relationship records
\begin{equation*}
r=(i,\,j,\,\text{type},\,\text{dir},\,s,\,c,\,\text{explicit},\,\text{quote}),
\end{equation*}
where $s\in[0,1]$ is the economic importance of the relationship \emph{to the
source firm}, $c\in[0,1]$ the agent's confidence that the relationship
currently exists, \emph{explicit} indicates whether the counterparty is named
in the text, and \emph{quote} is a verbatim evidence span ($\le$50 words).
Types cover supplier, customer, competitor, partner, licensor/licensee, input
dependency, and distribution channel. Crucially, agents are instructed to
resolve \emph{latent} mentions when context permits---mapping ``our largest
customer'' in a memory maker's filing to NVIDIA, or ``a limited number of
advanced foundry partners'' to TSMC---but must then set
$\text{explicit}=\text{false}$, cap confidence, and retain the anonymized
quote. This keeps the inference auditable and lets downstream users exclude
inferred edges entirely.

\textbf{Knowledge agents} receive no documents. Two independent instances
enumerate material bilateral relationships in the universe from parametric
knowledge alone, with calibrated confidences; the second is explicitly
prompted to be skeptical and to down-weight stale or terminated ties (e.g.,
Apple--Intel after the silicon transition). Their role is that of a prior:
they vote on edges the filings under-document, and their disagreement with
filing agents is itself a useful signal (\S\ref{sec:results-channels}).

\subsection{Entity resolution}
\label{sec:er}

Records refer to counterparties at three levels: universe tickers, external
companies (TSMC, OpenAI, McKesson), and non-entity categories (``third-party
payment processors'', ``uranium suppliers''). A curated alias map
canonicalizes the first two levels (Azure$\to$Microsoft, AWS$\to$Amazon,
Sam's Club$\to$Walmart, AmerisourceBergen$\to$Cencora); category mentions
cannot be resolved to nodes and are dropped from the graph but counted, since
their volume measures how much disclosed exposure remains unmeasurable at the
entity level. In our experiment 28.3\% of filing-agent records are such
category mentions---a direct quantification of the residual opacity of
disclosure text. External companies are retained as auxiliary nodes: they do
not appear in the final matrix but power the shared-counterparty layer.

\subsection{Fusion by weighted logit pooling}
\label{sec:fusion}

For an ordered pair $(i,j)$, let agent $a$ report confidence $p^{(a)}_{ij}$
with reliability weight $\omega_a$. We pool votes around a sparse prior
$\pi_0$:
\begin{equation}
p_{ij}=\sigma\!\Big(\operatorname{logit}\pi_0+\sum_a
\omega_a\big(\operatorname{logit}p^{(a)}_{ij}-\operatorname{logit}\pi_0\big)\Big),
\label{eq:pool}
\end{equation}
so that an agent reporting exactly the prior contributes nothing, agreement
compounds, and a lone low-reliability vote cannot push $p_{ij}$ far from
$\pi_0$. We set $\pi_0=0.02$ (an agnostic sparse-graph prior),
$\omega=1.0$ for explicit filing evidence, $0.6$ for inferred filing evidence,
$0.4$ for filing records whose retrieval failed, and $0.5$ for each knowledge
agent; within a channel, the strongest record casts the vote and duplicates
contribute at quarter weight. Dependency strength is the reliability-weighted
mean $s_{ij}$ of reported strengths, and the fused directed dependency is
$d_{i\to j}=p_{ij}\,s_{ij}$.

\subsection{Layers and economic stop-words}
\label{sec:idf}

The \textbf{direct layer} symmetrizes fused dependencies,
$W^{\mathrm{dir}}_{ij}=\max(d_{i\to j},d_{j\to i})$, retaining the directed
matrix for asymmetry analysis. The \textbf{shared-counterparty layer}
connects firms with common external dependencies. Let
$B_{ik}=d_{i\to k}$ for external entity $k$ and let $n_k$ be the number of
universe firms materially tied to $k$. We weight each shared counterparty by
its informativeness,
\begin{equation}
\mathrm{IDF}(k)=\log\frac{N}{1+n_k},\qquad
W^{\mathrm{sh}}_{ij}\propto\sum_k B_{ik}B_{jk}\,\mathrm{IDF}(k),
\label{eq:idf}
\end{equation}
the network analogue of removing stop-words: in our data TSMC ($n_k=18$),
Samsung ($14$), OpenAI ($11$), and SK~Hynix ($8$) receive the smallest
weights, while a shared dependence on, say, a specific flavor-ingredient
supplier remains informative. The final weighted matrix is a convex
combination $W=\alpha_1 \widetilde W^{\mathrm{dir}}+\alpha_2
\widetilde W^{\mathrm{sh}}$ with $(\alpha_1,\alpha_2)=(0.7,0.3)$ and each
layer max-normalized.

\subsection{Double-sided relative threshold}
\label{sec:threshold}

Binary adjacency uses per-node quantile thresholds. With $Q_i(\tau)$ the
$\tau$-quantile of firm $i$'s positive weights,
\begin{equation}
A_{ij}=\mathbf{1}\Big\{W_{ij}\ \ge\ \min\big(Q_i(\tau),Q_j(\tau)\big)\
\wedge\ W_{ij}\ge w_0\Big\},
\label{eq:thr}
\end{equation}
with $\tau=0.6$ and a small floor $w_0=0.04$. The $\min$ implements the
trade-network rule: an edge forms if $j$ is among $i$'s important partners
\emph{or} $i$ among $j$'s, preserving the one-sided relationships (small
supplier, dominant customer) that an absolute cutoff destroys. We show in
\S\ref{sec:results-abl} that a global threshold matched to the same edge
count isolates peripheral firms and concentrates degree on hubs.

\section{Experimental Setup}
\label{sec:setup}

\paragraph{Universe.} 42 Nasdaq-100 constituents spanning semiconductors
(NVDA, AMD, AVGO, QCOM, INTC, MU, ARM, MRVL, TXN, ADI, NXPI), semi equipment
(ASML, AMAT, LRCX, KLAC), EDA (CDNS, SNPS), hyperscalers and platforms (MSFT,
GOOGL, AMZN, META, AAPL, TSLA), software and internet (NFLX, PLTR, CRWV, DDOG,
CRWD, PANW, ADBE), utilities with data-center exposure (CEG, AEP, XEL),
consumer staples (PEP, COST, MDLZ, MNST, SBUX), and biotech (AMGN, GILD, VRTX,
REGN). The design embeds one dense expected ecosystem (AI infrastructure) and
two control groups expected to be nearly disconnected from it.

\paragraph{Documents.} Each firm's most recent 10-K or 20-F on SEC EDGAR
(fiscal 2025 for most firms; filing dates 2025-07 to 2026-05), retrieved and
read by the filing agents at experiment time. All evidence is therefore
point-in-time as of each firm's filing date. CoreWeave's first 10-K (filed
2026-03) is included, giving the pipeline a firm with essentially no history
in any curated database.

\paragraph{Agents.} Seven filing agents (six firms each) and two knowledge
agents, all instantiated from the same frontier LLM (Claude); an independent
adversarial audit agent with web search is used only for evaluation
(\S\ref{sec:results-precision}). Filing agents produced 547 records, of which
392 (71.7\%) resolved to entities (234 explicit, 148 inferred, 10 degraded
after failed retrieval); knowledge agents contributed 407 records. Fusion covers 601 ordered pairs over the 42
universe firms and 100 external entities.

\section{Results}
\label{sec:results}

\begin{table}[t]
\centering\small
\begin{tabular}{lrr}
\toprule
 & records & resolved \\
\midrule
Filing agents (42 filings) & 547 & 392 \\
\quad explicit (named in text) & & 234 \\
\quad inferred (latent mention) & & 148 \\
\quad degraded (failed retrieval) & & 10 \\
\quad category mention (dropped) & 155 & --- \\
Knowledge agent A & 188 & 187 \\
Knowledge agent B & 220 & 220 \\
\midrule
Ordered pairs with evidence & \multicolumn{2}{r}{601} \\
External entities & \multicolumn{2}{r}{100} \\
Final edges ($N{=}42$) & \multicolumn{2}{r}{149} \\
\bottomrule
\end{tabular}
\caption{Evidence collection and resolution statistics.}
\label{tab:stats}
\end{table}

\subsection{The measured network}

\begin{figure}[t]
\centering
\includegraphics[width=\columnwidth]{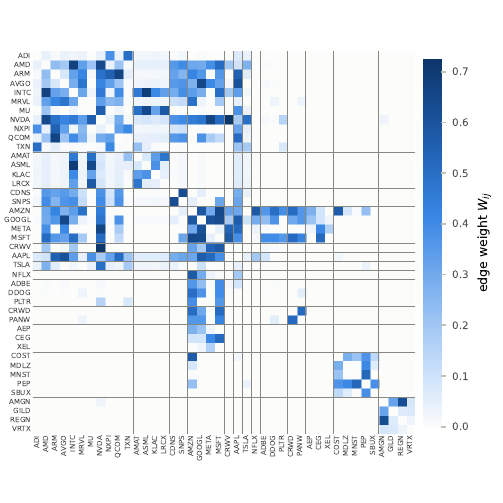}
\caption{Measured weighted adjacency matrix $W$, firms ordered by
fine-grained group (semis $\to$ equipment $\to$ EDA $\to$ hyperscalers $\to$
software $\to$ utilities $\to$ staples $\to$ biotech). Hairlines mark group
boundaries.}
\label{fig:heatmap}
\end{figure}

\begin{figure}[t]
\centering
\includegraphics[width=\columnwidth]{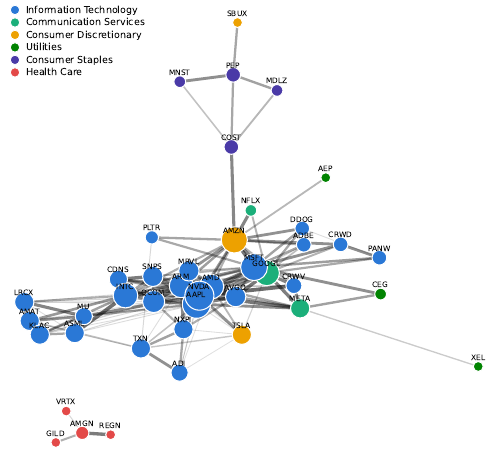}
\caption{Binary network $A$ (149 edges), node color = GICS sector, node size
$\propto$ degree, edge weight $\propto W_{ij}$. The AI-infrastructure core,
the staples cluster (attached via AMZN--COST), and the disconnected biotech
component emerge from text alone.}
\label{fig:network}
\end{figure}

The pipeline yields 149 edges (density 0.173), degree range 1--20 (median 6),
with no isolated firms. Figure~\ref{fig:heatmap} shows the weighted matrix
with firms ordered by fine-grained group: the AI-infrastructure block
(semis, equipment, EDA, hyperscalers, AI cloud) is dense and internally
structured, staples and biotech form their own small blocks, and
utility--hyperscaler edges appear exactly where power purchase agreements
exist. Figure~\ref{fig:network} shows the binary graph: one large component
organized around the NVDA--hyperscaler core, a staples cluster attached
through the AMZN--COST retail edge, and a separate biotech component---%
structure discovered entirely from disclosure text, with GICS labels used
only for coloring.

The highest-weight edges are economically exact: CRWV--NVDA ($W{=}0.73$;
CoreWeave's fleet is 100\% NVIDIA GPUs), ASML--INTC (0.69), AMD--INTC (0.67),
ARM--QCOM (0.66), META--NVDA (0.65), AVGO--GOOGL (0.65; custom TPU
co-design), ASML--MU (0.64), AMZN--MSFT (0.64), AMGN--REGN (0.63), CDNS--SNPS
(0.63).

\subsection{Precision: adversarial audit}
\label{sec:results-precision}

An independent audit agent with web access received a stratified sample of 30
edges (every 5th edge of the weight-sorted list) and was instructed to
\emph{refute} each: find evidence the relationship is absent, stale, or
immaterial as of 2025--2026. Results: 23 \textsc{confirmed}, 4
\textsc{plausible} (real but weakly documented), 3 \textsc{refuted}.
Precision is sharply weight-dependent (Table~\ref{tab:audit}): among sampled
edges with $W\ge 0.1$ (122 of 149 edges), 88\% are confirmed and none
refuted; all three refuted edges lie in the $W<0.1$ tail, and two of the
three are shared-counterparty-only edges with no direct evidence channel
(e.g., KLAC--QCOM, linked only through IDF-damped foundry exposure). The
weight itself is thus a usable reliability score, and a user wanting a
high-precision graph can simply raise $w_0$.

\begin{table}[t]
\centering\small
\begin{tabular}{lrrrr}
\toprule
weight bucket & $n$ & conf. & plaus. & refuted \\
\midrule
$W\ge 0.3$      & 21 & 20 & 1 & 0 \\
$0.1\le W<0.3$  & 4  & 2  & 2 & 0 \\
$W<0.1$         & 5  & 1  & 1 & 3 \\
\midrule
all             & 30 & 23 & 4 & 3 \\
\bottomrule
\end{tabular}
\caption{Adversarial audit of 30 stratified-sampled edges. Errors concentrate
entirely in the low-weight tail.}
\label{tab:audit}
\end{table}

\subsection{Agreement and disagreement with GICS}
\label{sec:results-gics}

The measured network should correlate with industry classification without
collapsing into it. 70.5\% of edges are within-GICS-sector against 37.3\%
expected under random placement (lift 1.9); at the finer 13-group level the
lift is 2.9. The interesting mass is the 44 cross-sector edges, which are not
noise but the pipeline's core product---relationships GICS is structurally
unable to represent (Table~\ref{tab:cross}). These include the
utility--hyperscaler power edges (CEG--MSFT from the Crane Clean Energy
Center PPA; CEG--META from the Clinton clean-energy agreement), AI-cloud
dependencies (CRWV--META, backed by a disclosed \$14.2B order), the
NVDA--TSLA compute relationship, the AMZN--COST retail edge, and the
PEP--SBUX ready-to-drink distribution partnership. A GICS-based factor model treats CEG as interchangeable with
regulated utilities like XEL; in the measured weighted matrix, by contrast,
\emph{all} of CEG's connectivity runs to the hyperscaler block---its only
edges are the two disclosed PPA counterparties, Microsoft and Meta.

\begin{table}[t]
\centering\small
\begin{tabular}{llr}
\toprule
edge & sectors & $W$ \\
\midrule
META--NVDA & Comm.\ $\times$ IT & 0.65 \\
AVGO--GOOGL & IT $\times$ Comm.\ & 0.65 \\
AMZN--MSFT & Cons.\ Disc.\ $\times$ IT & 0.64 \\
AMZN--COST & Cons.\ Disc.\ $\times$ Staples & 0.57 \\
CRWV--META & IT $\times$ Comm.\ & 0.54 \\
CEG--MSFT & Utilities $\times$ IT & 0.52 \\
NVDA--TSLA & IT $\times$ Cons.\ Disc.\ & 0.50 \\
CEG--META & Utilities $\times$ Comm.\ & 0.41 \\
\bottomrule
\end{tabular}
\caption{Selected cross-sector edges: economically real ties that no
single-label classification can encode.}
\label{tab:cross}
\end{table}

\subsection{Asymmetry of dependencies}
\label{sec:results-asym}

\begin{figure}[t]
\centering
\includegraphics[width=\columnwidth]{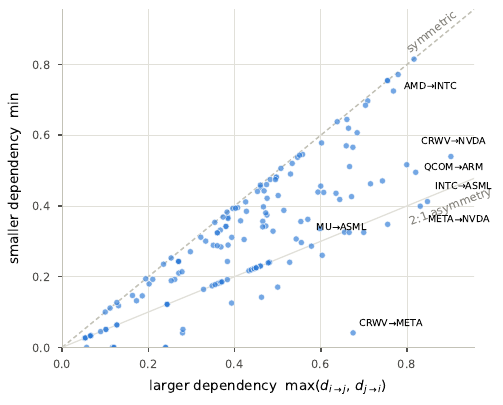}
\caption{Directed dependencies are asymmetric: each point is a firm pair,
plotting the larger against the smaller of $(d_{i\to j}, d_{j\to i})$.
Labels name the dependent firm (arrow points to the firm depended upon).}
\label{fig:asym}
\end{figure}

Because the pipeline measures $d_{i\to j}$ and $d_{j\to i}$ separately, it
recovers the size asymmetries that motivated relative thresholding
(Figure~\ref{fig:asym}). Among 137 firm pairs with a material dependency, the
mean relative asymmetry $|d_{i\to j}-d_{j\to i}|/\max(\cdot)$ is 0.27, and
12.4\% of pairs are more than 2:1 one-sided. The extremes are textbook cases:
CRWV$\to$META (0.67) vs.\ META$\to$CRWV (0.04)---a \$14B order that is
existential for CoreWeave and an implementation detail for Meta;
MU$\to$ASML (0.76) vs.\ ASML$\to$MU (0.35); QCOM$\to$ARM (0.82) vs.\
ARM$\to$QCOM (0.50). A symmetric similarity measure cannot represent this
information, and an absolute threshold would erase exactly these edges.

\subsection{Ablations}
\label{sec:results-abl}
\label{sec:results-channels}

\begin{table}[t]
\centering\small
\begin{tabular}{lrrrr}
\toprule
variant & edges & Jacc. & Gini & isol. \\
\midrule
full pipeline & 149 & --- & 0.40 & 0 \\
filing channel only & 109 & 0.61 & --- & 2 \\
knowledge channel only & 135 & 0.72 & --- & 0 \\
no IDF weighting & 153 & 0.92 & --- & 0 \\
absolute threshold$^\dagger$ & 149 & 0.69 & 0.45 & 1 \\
\bottomrule
\end{tabular}
\caption{Ablations. Jacc.\ = edge-set Jaccard similarity with the full
pipeline; Gini = degree concentration; isol.\ = isolated firms.
$^\dagger$Global threshold matched to the same edge count.}
\label{tab:abl}
\end{table}

\paragraph{Channels are complementary.} Filing-only and knowledge-only
networks overlap at Jaccard 0.435---each channel sees ties the other misses.
The filing channel contributes verbatim-documented edges the knowledge
channel underrates (e.g., the CEG PPAs, CRWV's disclosed concentration
percentages); the knowledge channel covers relationships filings leave
unnamed (42 firms' filings name essentially no equipment vendors, yet
MU--LRCX is real). 57.7\% of final edges are supported by both channels, and
41.6\% are anchored by at least one explicit named-entity quote.

\paragraph{IDF removes shared-hub artifacts.} Without
Eq.~\eqref{eq:idf}, the shared-counterparty layer manufactures edges between
firms whose only commonality is a ubiquitous counterparty: AMD--QCOM,
ASML--QCOM, AAPL--CDNS, GOOGL--INTC, and MU--TSLA all enter the no-IDF graph
via shared TSMC/Samsung exposure. These are precisely the ``uses
semiconductors''-grade non-edges the senator-network analysis warns against.

\paragraph{Relative thresholds protect the periphery.} A global threshold
matched to the same 149 edges reallocates them toward hubs: degree Gini rises
from 0.40 to 0.45, the maximum degree rises, VRTX becomes isolated, and
one-sided but real edges (AAPL--ADI's residual supplier tie, AMAT--ASML,
ARM--TSLA's architecture licensing) are destroyed. The double-sided
rule keeps every firm attached through its own most important relationships,
which is the property a downstream community-detection or spillover model
needs.

\section{Discussion and Limitations}
\label{sec:limitations}

\paragraph{Point-in-time discipline.} Filing-channel evidence is point-in-time
by construction (each record carries its filing date), but knowledge-channel
evidence inherits the LLM's training cutoff and is \emph{not} reconstructible
for historical dates: a 2021 backtest cannot use a 2026 model's priors
without look-ahead. Historical applications should either restrict to the
filing channel---which our ablation shows sacrifices recall (109 vs.\ 149
edges) but not precision---or use models with controlled cutoffs.

\paragraph{Retrieval fragility.} Large filings are read through an extraction
interface that truncates; several concentration disclosures were recovered
only from XBRL exhibit tables, and 1.8\% of records fell back to
knowledge-grade confidence after failed retrieval. The 28.3\% of records
dropped as category mentions bound what entity-level measurement can extract
from disclosure alone; matching category mentions to entities via structured
data (customs records, input--output tables) is the natural extension.

\paragraph{Inferred edges carry model risk.} 41\% of resolved filing records
involve latent-mention inference. Our audit shows errors concentrate in
low-weight, single-layer edges rather than in inferences per se (the
MU$\to$NVDA resolution of ``one customer, 17\% of revenue'' is confirmed by
subsequent disclosure), but every inferred edge retains its flag, quote, and
confidence so users can apply their own tolerance.

\paragraph{Scope.} We measure one snapshot of 42 firms; the design scales
linearly in firms and time (each filing is processed independently), and the
companion work applies dynamic community detection and cross-entity signal
propagation to the resulting matrix sequence. We deliberately make no return-%
predictability claims here: the contribution is the measurement instrument,
audited on its own terms.

\section{Conclusion}

We posed adjacency-matrix construction as an LLM measurement problem and
showed that a multi-agent pipeline with entity resolution, logit-pooled
fusion, ubiquity filtering, and relative thresholding produces a firm network
that is precise where it claims to be (88\% audited precision on
$W\ge0.1$ edges, errors confined to the flagged low-weight tail), agrees with
industry classification exactly as much as it should, and recovers the
cross-sector dependencies---power purchase agreements, GPU supply chains,
retail channels---that make firm networks worth measuring. The released
records, with verbatim evidence and explicit/inferred flags, are intended as
a reusable substrate for community detection, spillover estimation, and
graph-based asset pricing.

\section*{Acknowledgments}
Filing text was obtained from SEC EDGAR. All extraction, fusion, and audit
code accompanies the paper.

\bibliography{references}

@article{cohen2008economic,
  title={Economic Links and Predictable Returns},
  author={Cohen, Lauren and Frazzini, Andrea},
  journal={The Journal of Finance},
  volume={63}, number={4}, pages={1977--2011}, year={2008}
}

@article{menzly2010market,
  title={Market Segmentation and Cross-predictability of Returns},
  author={Menzly, Lior and Ozbas, Oguzhan},
  journal={The Journal of Finance},
  volume={65}, number={4}, pages={1555--1580}, year={2010}
}

@article{ali2020shared,
  title={Shared Analyst Coverage: Unifying Momentum Spillover Effects},
  author={Ali, Usman and Hirshleifer, David},
  journal={Journal of Financial Economics},
  volume={136}, number={3}, pages={649--675}, year={2020}
}

@article{hoberg2010product,
  title={Product Market Synergies and Competition in Mergers and Acquisitions: A Text-Based Analysis},
  author={Hoberg, Gerard and Phillips, Gordon},
  journal={The Review of Financial Studies},
  volume={23}, number={10}, pages={3773--3811}, year={2010}
}

@article{hoberg2016text,
  title={Text-Based Network Industries and Endogenous Product Differentiation},
  author={Hoberg, Gerard and Phillips, Gordon},
  journal={Journal of Political Economy},
  volume={124}, number={5}, pages={1423--1465}, year={2016}
}

@article{lee2015search,
  title={Search-Based Peer Firms: Aggregating Investor Perceptions through Internet Co-Searches},
  author={Lee, Charles M.~C. and Ma, Paul and Wang, Charles C.~Y.},
  journal={Journal of Financial Economics},
  volume={116}, number={2}, pages={410--431}, year={2015}
}

@article{kaustia2021common,
  title={Common Analysts: Method for Defining Peer Firms},
  author={Kaustia, Markku and Rantala, Ville},
  journal={Journal of Financial and Quantitative Analysis},
  volume={56}, number={5}, pages={1505--1536}, year={2021}
}

@article{schwenkler2019network,
  title={The Network of Firms Implied by the News},
  author={Schwenkler, Gustavo and Zheng, Hannan},
  journal={SSRN Working Paper No.\ 3320859},
  year={2019}
}

@article{wichmann2020extracting,
  title={Extracting Supply Chain Maps from News Articles Using Deep Neural Networks},
  author={Wichmann, Pascal and Brintrup, Alexandra and Baker, Simon and Woodall, Philip and McFarlane, Duncan},
  journal={International Journal of Production Research},
  volume={58}, number={17}, pages={5320--5336}, year={2020}
}

@article{liu2024supply,
  title={Supply Chain Network Extraction and Entity Classification Leveraging Large Language Models},
  author={Liu, Tong and Meidani, Hadi},
  journal={arXiv preprint arXiv:2410.13051},
  year={2024}
}

@article{almahri2024enhancing,
  title={Enhancing Supply Chain Visibility with Knowledge Graphs and Large Language Models},
  author={AlMahri, Sara and Xu, Liming and Brintrup, Alexandra},
  journal={International Journal of Production Research},
  year={2025},
  note={arXiv:2408.07705}
}

@article{jackson2025supply,
  title={Supply Chain Mapping through Retrieval-Augmented Generation: Applications to the Electronics Industry},
  author={Jackson, Ilya and Saenz, Maria Jesus and Ma, Bill Jiaqi and Ivanov, Dmitry},
  journal={Journal of the Operational Research Society},
  year={2025}
}

@article{culot2023using,
  title={Using Supply Chain Databases in Academic Research: A Methodological Critique},
  author={Culot, Giovanna and Podrecca, Matteo and Nassimbeni, Guido and Orzes, Guido and Sartor, Marco},
  journal={Journal of Supply Chain Management},
  volume={59}, number={1}, pages={3--25}, year={2023}
}

@inproceedings{sharma2022finred,
  title={{FinRED}: A Dataset for Relation Extraction in Financial Domain},
  author={Sharma, Soumya and Nayak, Tapas and Bose, Arusarka and Meena, Ajay Kumar and Dasgupta, Koustuv and Ganguly, Niloy and Goyal, Pawan},
  booktitle={Companion Proceedings of the Web Conference 2022 (FinWeb)},
  pages={595--597}, year={2022}
}

@inproceedings{kaur2023refind,
  title={{REFinD}: Relation Extraction Financial Dataset},
  author={Kaur, Simerjot and Smiley, Charese and Gupta, Akshat and Sain, Joy and Wang, Dongsheng and Siddagangappa, Suchetha and Aguda, Toyin and Shah, Sameena},
  booktitle={Proceedings of the 46th International ACM SIGIR Conference on Research and Development in Information Retrieval},
  year={2023}
}

@inproceedings{rajpoot2023gptfinre,
  title={{GPT-FinRE}: In-context Learning for Financial Relation Extraction using Large Language Models},
  author={Rajpoot, Pawan Kumar and Parikh, Ankur},
  booktitle={Proceedings of the Sixth Workshop on Financial Technology and Natural Language Processing (FinNLP)},
  year={2023}
}

@inproceedings{li2024findkg,
  title={{FinDKG}: Dynamic Knowledge Graphs with Large Language Models for Detecting Global Trends in Financial Markets},
  author={Li, Xiaohui Victor},
  booktitle={Proceedings of the 5th ACM International Conference on AI in Finance (ICAIF)},
  year={2024}
}

@inproceedings{arun2025finreflectkg,
  title={{FinReflectKG}: Agentic Construction and Evaluation of Financial Knowledge Graphs},
  author={Arun, Abhinav and Dimino, Fabrizio and Agarwal, Tejas Prakash and Sarmah, Bhaskarjit and Pasquali, Stefano},
  booktitle={Proceedings of the 6th ACM International Conference on AI in Finance (ICAIF)},
  year={2025}
}

@inproceedings{chung2023modeling,
  title={Modeling Momentum Spillover with Economic Links Discovered from Financial Documents},
  author={Chung, Alex and Tanaka-Ishii, Kumiko},
  booktitle={Proceedings of the 4th ACM International Conference on AI in Finance (ICAIF)},
  year={2023}
}

@article{vamvourellis2023company,
  title={Company Similarity Using Large Language Models},
  author={Vamvourellis, Dimitrios and Toth, M{\'a}t{\'e} and Bhagat, Snigdha and Desai, Dhruv and Mehta, Dhagash and Pasquali, Stefano},
  journal={arXiv preprint arXiv:2308.08031},
  year={2023}
}

@inproceedings{chen2018incorporating,
  title={Incorporating Corporation Relationship via Graph Convolutional Neural Networks for Stock Price Prediction},
  author={Chen, Yingmei and Wei, Zhongyu and Huang, Xuanjing},
  booktitle={Proceedings of the 27th ACM International Conference on Information and Knowledge Management (CIKM)},
  year={2018}
}

@article{feng2019temporal,
  title={Temporal Relational Ranking for Stock Prediction},
  author={Feng, Fuli and He, Xiangnan and Wang, Xiang and Luo, Cheng and Liu, Yiqun and Chua, Tat-Seng},
  journal={ACM Transactions on Information Systems},
  volume={37}, number={2}, pages={1--30}, year={2019}
}

@article{kim2019hats,
  title={{HATS}: A Hierarchical Graph Attention Network for Stock Movement Prediction},
  author={Kim, Raehyun and So, Chan Ho and Jeong, Minbyul and Lee, Sanghoon and Kim, Jinkyu and Kang, Jaewoo},
  journal={arXiv preprint arXiv:1908.07999},
  year={2019}
}

@article{chen2023chatgpt,
  title={{ChatGPT} Informed Graph Neural Network for Stock Movement Prediction},
  author={Chen, Zihan and Zheng, Lei and Lu, Cheng and Yuan, Jialu and Zhu, Di},
  journal={arXiv preprint arXiv:2306.03763},
  year={2023}
}

@article{holland1983stochastic,
  title={Stochastic Blockmodels: First Steps},
  author={Holland, Paul W. and Laskey, Kathryn Blackmond and Leinhardt, Samuel},
  journal={Social Networks},
  volume={5}, number={2}, pages={109--137}, year={1983}
}

@article{chen2018network,
  title={Network Cross-Validation for Determining the Number of Communities in Network Data},
  author={Chen, Kehui and Lei, Jing},
  journal={Journal of the American Statistical Association},
  volume={113}, number={521}, pages={241--251}, year={2018}
}

@article{li2020network,
  title={Network Cross-Validation by Edge Sampling},
  author={Li, Tianxi and Levina, Elizaveta and Zhu, Ji},
  journal={Biometrika},
  volume={107}, number={2}, pages={257--276}, year={2020}
}

@article{wang2017likelihood,
  title={Likelihood-Based Model Selection for Stochastic Block Models},
  author={Wang, Y.~X.~Rachel and Bickel, Peter J.},
  journal={The Annals of Statistics},
  volume={45}, number={2}, pages={500--528}, year={2017}
}

\appendix

\section{Reproducibility Details}
\label{app:repro}

\paragraph{Hyperparameters.} Sparse prior $\pi_0=0.02$; channel weights
$\omega$: explicit filing 1.0, inferred filing 0.6, failed-retrieval filing
0.4, knowledge agents 0.5 each; duplicate votes within a channel 0.25$\times$;
layer mix $(\alpha_1,\alpha_2)=(0.7,0.3)$; node quantile $\tau=0.6$; floor
$w_0=0.04$. Confidences are clamped to $[0.05,0.98]$ before pooling.

\paragraph{Record schema.} Each filing record: source ticker, resolved
target, relation type (8 values), explicit flag, inferred flag, strength,
confidence, verbatim quote ($\le$50 words), document id, filing date. Each
knowledge record: entity pair, relation type (7 values), strength,
confidence, one-line justification.

\paragraph{Audit protocol.} Stratified sample of 30 edges (every 5th of the
weight-sorted 149); auditor instructed to refute; verdicts
\textsc{confirmed}/\textsc{plausible}/\textsc{refuted} with a source URL per
verdict. The auditor shares no state with extraction agents.

\section{Economic Stop-Words}
\label{app:stopwords}

Most-connected external entities and their IDF weights (normalized):
TSMC ($n_k{=}18$, 0.21), Samsung (14, 0.28), OpenAI (11, 0.34), SK Hynix
(8, 0.41), Walmart (5, 0.52), ASE Group, Dell, Cencora, Foxconn (4, 0.57),
Cisco, Sony, STMicroelectronics, McKesson, Coca-Cola, Arrow Electronics
(3, 0.63). Rare shared counterparties (specialty suppliers, single
collaborators) retain weight near 1.0.

\end{document}